\documentclass[preprintnumbers,amsmath,amssymb,showpacs]{revtex4}

\usepackage[                   
            pdfstartview=FitH,
            colorlinks, 
            pdfborder=001,   
            linkcolor=blue,
            anchorcolor=blue,
            citecolor=blue,
            urlcolor=blue
            ]{hyperref}
\usepackage{graphicx}
\usepackage{dcolumn}
\usepackage{bm}
\usepackage{subfigure}

\begin{document}
\title{A $(k+1)$-partite entanglement measure of $N$-partite quantum states}

\author{Yan Hong}
 \affiliation {School of Mathematics and Science, Hebei GEO University, Shijiazhuang 050031,  China}

\author{Xianfei Qi}
 \affiliation {School of Mathematics and Statistics, Shangqiu Normal University, Shangqiu 476000,  China}

\author{Ting Gao}
\email{gaoting@hebtu.edu.cn} \affiliation {School of Mathematical Sciences, Hebei Normal University, Shijiazhuang 050024,  China}

\author{Fengli Yan}
\email{flyan@hebtu.edu.cn} \affiliation {College of Physics, Hebei Key Laboratory of Photophysics Research and Application,
Hebei Normal University, Shijiazhuang 050024,  China}

\begin{abstract}
The concept of \textquotedblleft the permutationally invariant part of a density matrx\textquotedblright constitutes an important tool for entanglement characterization of multiqubit systems. In this paper, we first present $(k+1)$-partite entanglement measure of $N$-partite quantum system,  which possesses desirable properties of an entanglement measure. Moreover, we give strong bounds on this measure by considering the permutationally invariant part of a multipartite state. We give two definitions of efficient measurable degree of $(k+1)$-partite entanglement. Finally, several concrete examples are given to illustrate the effectiveness of our results.
\end{abstract}


\maketitle

\section{Introduction}
Quantum entanglement is one of the most distinctive features of quantum mechanics as compared to classical theory \cite{RMP81.865}. It is also recognized as a remarkable resource in the various areas of quantum information processing such as quantum computation \cite{Nature2000}, quantum teleportation \cite{PRL70.1895,EPL84.50001} and dense coding \cite{PRL69.2881}.  Therefore, measuring and detection of entanglement are fundamental tasks in the theory of quantum entanglement \cite{QIC2007,PR474.1,JPA2014}.

For bipartite quantum systems, quantum states consist of separable states and entangled states. In recent years, much effort has been devoted to distinguish separable from entangled states in both experiment and theory  \cite{TCS2002,AMP2010,FP2013,NRP2019}. However, the classification of multipartite quantum states becomes much more complex because multipartite quantum systems contain more than two individual subsystems. For example, the $N$-partite quantum states can be divided into $k$-separable states and $k$-nonseparable states ($2\leqslant k\leqslant N$) \cite{EPL104.20007}. The detection of $k$-nonseparability has been investigated extensively, many efficient criteria \cite{QIC2008,QIC2010,PRA82.062113,EPL104.20007,PRA91.042313,SR5.13138,PRA93.042310,SciChina2017,CPB2018,QIP2020} and computable measures \cite{PRA68.042307,PRL93.230501,PRA83.062325,PRA86.062323,PRL112.180501} have been presented. The $N$-partite quantum states can also be divided into $k$-producible states $(1\leq k\leq N-1)$ and $(k+1)$-partite entangled states \cite{PLA2021}. It is worth noting that the $(k+1)$-partite entanglement and the $k$-nonseparability are two different concepts involving the partitions of subsystem in $N$-partite quantum systems, and they are equivalent only in some special cases.

In this work, we  introduce a quantitative measure of $(k+1)$-partite entanglement for general multipartite quantum states and prove explicitly  that this measure satisfies many properties of multipartite entanglement measures. Moreover, we obtain strong bounds on this measure by considering the permutationally invariant part of a multipartite state. Furthermore, we show that if the permutationally invariant part of a state is ($k$+1)-partite entangled, then so is the actual state.  We define also two efficient numbers based on some simple and powerful $(k+1)$-partite entanglement criteria of $N$ qubit states. In addition, we discuss the application of these two quantities in several examples.

\section{Preliminaries}

 For an $N$-partite quantum system with Hilbert space  $\mathcal{H}_1\otimes\mathcal{H}_2\otimes\cdots\otimes\mathcal{H}_N$, a pure state $|\psi\rangle$ is  $k$-producible ( $1\leq k\leq N-1$), if it can be represented as
$|\psi\rangle=\bigotimes\limits_{t=1}^m|\psi_t\rangle_{A_t}$
under the  partition $A=A_1|A_2|\cdots |A_m$ satisfying
\begin{eqnarray}\label{partition}
\bigcup\limits_{t=1}^{m}A_t=\{1,2,\cdots, N\}, A_t\cap A_{t'} =\varnothing \textrm{ for any }  t\neq t',
\end{eqnarray}
with the number of particles $|A_t|$ in the subset $A_t$ being no more than $k$ for any $t$, and the substate $|\psi_t\rangle_{A_t}\in\bigotimes\limits_{i\in A_t}\mathcal{H}_i$ \cite{PLA2021}.
For the  $N$-partite mixed state $\rho$, if it can be written as a convex combination of $k$-producible pure states,
i.e., $\rho=\sum\limits_ip_i|\psi_i\rangle\langle\psi_i|$,
 then it is called  $k$-producible,   where the pure state $|\psi_i\rangle$ might be $k$-producible in different partitions satisfying (\ref{partition}). If a quantum state is not $k$-producible,  it contains $(k+1)$-partite entanglement.

For an $N$ qubit quantum state $\rho$  in the Hilbert space  $\mathcal{H}_1\otimes\mathcal{H}_2\otimes\cdots\otimes\mathcal{H}_N$,  its permutationly invariant (PI) part is defined as
\begin{equation}\label{}\nonumber
\rho^{\textrm{PI}}=\dfrac{1}{N!}\sum\limits_{i=1}^{N!}\Pi_i\rho\Pi_i^\dag,
\end{equation}
where the sum takes all over $N!$ permutations $\{\Pi_i\}$ of  $N$ particles.
Let $U:=U_1\otimes U_2\otimes\cdots \otimes U_N$ be the local unitary transformation with unitary operator $U_i$ acting on $i$-th subsystem, and $\rho_U^{(\textrm{PI})}:=\dfrac{1}{N!}\sum\limits_{i=1}^{N!}\Pi_iU\rho U^\dag\Pi_i^\dag$.

\section{A measure of $(k+1)$-partite entanglement}

 Let us now introduce a $(k+1)$-partite entanglement measure called $(k+1)$-PE concurrence for quantum states in  $N$-partite  Hilbert space  $\mathcal{H}_1\otimes\mathcal{H}_2\otimes\cdots\otimes\mathcal{H}_N$.
 For an $N$-partite pure state $|\psi\rangle\in\mathcal{H}_1\otimes
\mathcal{H}_2\otimes\cdots\otimes\mathcal{H}_N$,  we define the $(k+1)$-PE concurrence as
\begin{equation}\label{}\nonumber
E_{k}(|\psi\rangle)=\min\limits_{A}\dfrac{\sum\limits_{t=1}^{m}\sqrt{2\left[1-\textrm{Tr}(\rho^2_{A_t})\right]}}{m},
\end{equation}
where $\rho_{A_t}$ is the reduced density operator of pure state $|\psi\rangle$ for subsystem $A_t$, and the minimum is taken over all possible partitions satisfying (\ref{partition}).

For the $N$-partite mixed state $\rho$ in  the Hilbert space  $\mathcal{H}_1\otimes\mathcal{H}_2\otimes\cdots\otimes\mathcal{H}_N$, we define the $(k+1)$-PE concurrence  as
\begin{equation}\label{kmix}\nonumber
E_{k}(\rho)=\inf\limits_{\{ p_i,|\psi_i\rangle\}}\sum\limits_ip_iE_{k}(|\psi_i\rangle),
\end{equation}
with the infimum  running over all possible pure state ensemble decompositions $\{p_i,|\psi_i\rangle\}$ of the mixed state $\rho$.

The  $(k+1)$-PE concurrence $E_{k}(\rho)$ is multipartite entanglement measure  satisfying  properties:

(P1) vanishing on  all $k$-producible states,  $E_{k}(\rho)=0$ for any $k$-producible state $\rho$.

(P2)  $E_{k}(\rho)>0$ for any quantum states $\rho$ containing $(k+1)$-partite entanglement.

(P3) invariant under local unitary transformations, $E_{k}(U_1\otimes\cdots\otimes U_N\rho U_1^\dag\otimes\cdots\otimes U_N^\dag)= E_{k}(\rho)$.

(P4) convexity, $E_{k}(\sum\limits_ip_i\rho_i)\leq \sum\limits_ip_iE_{k}(\rho_i)$.

(P5) nonincreasing under local operation and classical communication,   $E_{k}(\Lambda_{\textrm{LOCC}}(\rho))\leq E_{k}(\rho)$.

(P6) subadditivity, $E_{k}(\rho\otimes\sigma)\leq E_{k}(\rho)+E_{k}(\sigma).$

The details of proof is in the Appendix A.  Moreover, we can demonstrate the following result.

$\emph{Proposition 1}. $ For any $N$ qubit quantum state $\rho$,
  a lower bound of $(k+1)$-PE concurrence $E_{k}(\rho)$  can be expressed in terms of PI part of $\rho$ as follows
\begin{equation}\label{EPI}
E_{k}(\rho)\geq\max E_{k}(\rho_U^{(\textrm{PI})}),
\end{equation}
where the maximum takes over all possible local unitary transformations.

\emph{Proof}: To derive the above result, let $A_1|\cdots|A_m$ be a partition satisfying (\ref{partition}), then  $\Pi_i(A_1)|\cdots|\Pi_i(A_m)$ is also a partition satisfying (\ref{partition}). For any pure state $|\psi\rangle$, $\Pi_i|\psi\rangle$ is still pure state, so
 we can obtain
\begin{equation}\label{EPpure}
E_{k}(|\psi\rangle)=E_{k}(\Pi_i|\psi\rangle),
\end{equation}
by the definition of  $(k+1)$-PE concurrence for pure states.

The PI part of pure state $\rho=|\psi\rangle\langle\psi|$ is
$\rho^{\textrm{PI}}=\dfrac{1}{N!}\sum\limits_{i=1}^{N!}\Pi_i|\psi\rangle\langle\psi|\Pi_i^\dag,$
then one has
\begin{equation}\label{EPIpure}
\begin{array}{rl}
E_{k}(\rho^{\textrm{PI}})
\leq\dfrac{1}{N!}\sum\limits_{i=1}^{N!}E_{k}(\Pi_i|\psi\rangle)=\dfrac{1}{N!}\sum\limits_{i=1}^{N!}E_{k}(|\psi\rangle)=E_{k}(|\psi\rangle),
\end{array}
\end{equation}
where the inequality holds by the convexity of $(k+1)$-PE concurrence, the equality follows from (\ref{EPpure}).
By the  definition of  $(k+1)$-PE concurrence for mixed states, the  convexity of $(k+1)$-PE concurrence and inequality (\ref{EPIpure}), we can get
\begin{equation}\label{}\nonumber
\begin{array}{rl}
E_{k}(\rho^{\textrm{PI}})
\leq E_{k}(\rho)
\end{array}
\end{equation}
 for   mixed state $\rho$. Since the $(k+1)$-PE concurrence remains unchanged under any local unitary transformation,
we have \begin{equation}\label{}\nonumber
\max E_{k}(\rho_U^{(\textrm{PI})})\leq E_{k}(\rho),
\end{equation}
where the maximum takes over all possible local unitary transformations $U=U_1\otimes U_2\otimes\cdots \otimes U_N$. That's what we're trying to prove.

By the items (P1) and (P2) of  entanglement measure $(k+1)$-PE concurrence and Proposition 1, for any $k$-producible state $\rho$, we can get $E_k(\rho^{(\textrm{PI})})=0$, this  means that PI part $\rho^{(\textrm{PI})}$ is also $k$-producible.  All in all, we get the following conclusion as a $(k+1)$-partite entanglement criterion.

$\emph{Corollary}.$  If  $N$ qubit quantum state  $\rho$  is $k$-producible,   PI part $\rho^{(\textrm{PI})}$ is also $k$-producible. That is, if PI part $\rho^{(\textrm{PI})}$ of quantum state $\rho$ contains $(k+1)$-partite entanglement, the original quantum state $\rho$ also contains $(k+1)$-partite entanglement.

\section{The degree of $(k+1)$-partite entanglement }

There are some criteria for identifying  $(k+1)$-partite entanglement t such as the methods mentioned in Ref. \cite{PLA2021}:
 If $N$ qubit quantum state $\rho$ is a $k$-producible, then it must satisfy
\begin{equation}\label{PLAk-p1}
\begin{array}{rl}
(2^r-2)|\rho_{1,2^N}|\leq\sum\limits_{i=2}^{2^N-1}\sqrt{\rho_{i,i}\rho_{2^N-i+1,2^N-i+1}}
\end{array}
\end{equation}
  if let $|\phi_1\rangle=|0\rangle^{\otimes N}$, $|\phi_2\rangle=|1\rangle^{\otimes N}$ for Theorem 1 of  Ref. \cite{PLA2021}, where $r=\frac{N}{k}$ for $k|N$,  $r=[\frac{N}{k}]$ for $k\nmid N$, and
\begin{equation}\label{PLAk-p2}
\begin{array}{rl}
\sum\limits_{0\leq i\neq j\leq N-1}|\rho_{2^i+1,2^j+1}|\leq\sum\limits_{0\leq i\neq j\leq N-1}\sqrt{\rho_{1,1}\rho_{2^i+2^j+1,2^i+2^j+1}}
+(k-1)\sum\limits_{i=0}^{N-1}|\rho_{2^i+1,2^i+1}|
\end{array}
\end{equation}
if let $\{\omega_1,\cdots,\omega_T\}=\{1\}$, $|\psi_i^s\rangle=|0\rangle^{\otimes (i-1)}|1\rangle|0\rangle^{\otimes (N-i)}$ for Theorem 2 of  Ref. \cite{PLA2021}.
According to inequality (\ref{PLAk-p1}) and inequality (\ref{PLAk-p2}), we can introduce the following two quantities, both of which reflect the degree of $(k+1)$-partite entanglement, namely
\begin{equation}\label{D1}
\begin{array}{rl}
D_k(N)=\log_2\Big(\dfrac{B}{A}+2\Big),
\end{array}
\end{equation}
\begin{equation}\label{D2}
\begin{array}{rl}
\widetilde{D}_k(N)=\frac{C-D}{E}+1,
\end{array}
\end{equation}
where
\begin{equation}\label{}\nonumber
\begin{array}{rl}
A=&|\rho_{1,2^N}|,\\
B=&\sum\limits_{i=2}^{2^N-1}\sqrt{\rho_{i,i}\rho_{2^N-i+1,2^N-i+1}},\\
C=&\sum\limits_{0\leq i\neq j\leq N-1}|\rho_{2^i+1,2^j+1}|,\\
D=&\sum\limits_{0\leq i\neq j\leq N-1}\sqrt{\rho_{1,1}\rho_{2^i+2^j+1,2^i+2^j+1}},\\
E=&\sum\limits_{i=0}^{N-1}|\rho_{2^i+1,2^i+1}|.
\end{array}
\end{equation}
If $D_k(N)<\lceil\frac{N}{k}\rceil$ or $\widetilde{D}_k(N)>k$, then PI part of quantum state $\rho$ contains $(k+1)$-partite entanglement. Now we will give two examples.

$\emph{Example 1}.$ For an $N$ qubit quantum state $\rho(p)=p|G_N\rangle\langle G_N|+\dfrac{1-p}{2^N}\textbf{I}$ with $|G_N\rangle=\dfrac{|0\rangle^{\otimes N}+|1\rangle^{\otimes N}}{\sqrt{2}}$, we have
\begin{equation}\label{}\nonumber
\begin{array}{rl}
D_k(N,p)=\log_2\Big(\dfrac{2^{N-1}+p-1}{2^{N-2}p}\Big).
\end{array}
\end{equation}
Clearly,  $\rho(p)$ is a permutationly invariant quantum state.
 The  quantum state $\rho$ contains  $(k+1)$-partite entanglement when $D_k(N)<\lceil\frac{N}{k}\rceil$. It can be seen from FIG. 1 that the larger $N$ is, the larger the value range of $p$ corresponding to  $\rho(p)$ containing $2$-partite entanglement is.
Using $D_{1}(N,p)<N$, one has $p>\dfrac{1-2^{N-1}}{1-2^{2N-2}}$. That is,  $\rho(p)$ contains $2$-partite entanglement when
$\dfrac{1-2^{N-1}}{1-2^{2N-2}}<p\leq1$.
Since $\lim\limits_{N\rightarrow+\infty}\dfrac{1-2^{N-1}}{1-2^{2N-2}}=0$,  $\rho(p)$ can be arbitrarily close to white noise when $N$ increases, but it still contains $2$-partite entanglement.
\begin{figure}
\begin{center}
{\includegraphics[scale=0.5]{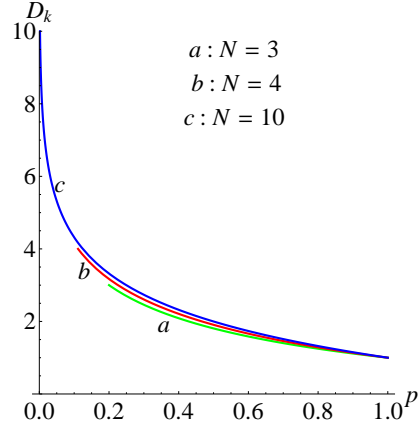}} \caption[Illustration of
]{(Color online) For an $N$ qubit quantum state
$\rho(p)=p|G_N\rangle\langle G_N|+\dfrac{1-p}{2^N}\textbf{I}$ , this figure  shows  $D_k(N,p)$ when $N$=3, 4, and 10, that is,
the green, red, and blue lines represent the graphs of $D_k(3,p), D_k(4,p),$ and $D_k(10,p)$, respectively. It should be noted that the part of the graphs corresponding to $D_k(N,p)\geq N$ is removed.}
\end{center}
\end{figure}

$\emph{Example 2}.$ Consider the $N$ qubit  mixture of $|W_N\rangle=\dfrac{|10\cdots0\rangle+|01\cdots0\rangle+\cdots+|00\cdots1\rangle}{\sqrt{N}}$ and white noise, which is  $\rho(p)=(1-p)|W_N\rangle\langle W_N|+\dfrac{p}{2^N}\textbf{I}$. After calculation, we get
\begin{equation}\label{}\nonumber
\begin{array}{rl}
\widetilde{D}_k(N,p)=\dfrac{N2^N-(N2^N+N^2-2N)p}{2^N-(2^N-N)p}.
\end{array}
\end{equation}
Obviously,  $\rho(p)$ is a permutationly invariant quantum state. The $\rho(p)$ contains  $(k+1)$-partite entanglement when $\widetilde{D}_k(N)>k$.  The $\rho(p)$ contains $N$-partite entanglement when $0\leq p<\dfrac{2^N}{2^N+2N^2-3N}$. The FIG. 2 shows that the larger $N$ is, the larger the value range of $p$ corresponding to $\rho(p)$  containing $N$-partite entanglement is.   In fact, by $\lim\limits_{N\rightarrow+\infty}\dfrac{2^N}{2^N+2N^2-3N}=1$, we can conclude that a large value of $p$  can ensure that $\rho(p)$ contains $N$-partite entanglement  as $N$ increases.

\begin{figure}
\begin{center}
{\includegraphics[scale=0.5]{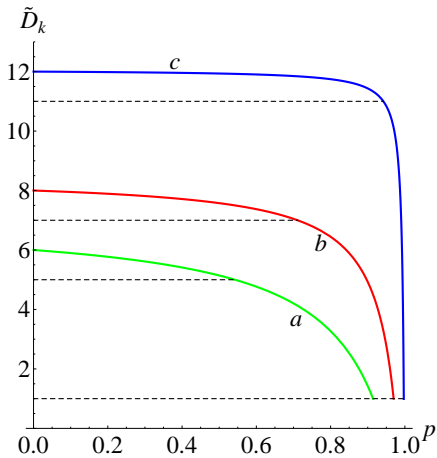}} \caption[Illustration of
]{(Color online) For an $N$ qubit quantum state
$\rho(p)=(1-p)|W_N\rangle\langle W_N|+\dfrac{p}{2^N}\textbf{I}$ ,
the green, red, and blue lines represent the graphs of $\widetilde{D}_k(6,p), \widetilde{D}_k(8,p),$ and $\widetilde{D}_k(12,p)$, respectively. The part of the graphs corresponding to $\widetilde{D}_k(N,p)<1$ is not drawn.}
\end{center}
\end{figure}

\section{Conclusions}
In conclusion, we presented a measure of $(k+1)$-partite entanglement for general multipartite quantum states. The rigorous proof shows that it satisfies the requirements for a proper multipartite entanglement measure. We have derived bounds on this measure by considering the PI part of a multipartite state. As an immediate consequence of these bounds, we showed that whenever the PI part of a state is $(k+1)$-partite entangled, then so is the state itself. Based on some $(k+1)$-partite entanglement criteria of $N$-qubit states, we defined two numbers that can be efficiently measurable and quantify the degree of separability of the state. The results in this paper may shed some new light on the study of entanglement properties of multipartite quantum systems.

\begin{center}
{\bf ACKNOWLEDGMENTS}
\end{center}

This work was supported by  the National Natural Science Foundation of China under Grant Nos. 12071110, 62271189, the Hebei Natural Science Foundation of China under Grant No. A2020205014, funded by Science and Technology Project of Hebei Education Department under Grant Nos. ZD2020167, ZD2021066, supported by PhD Research Startup Foundation of Hebei GEO University (Grant BQ201615) and the Key Scientific Research Project of Henan Higher Education Institutions under Grant No.22B140006.

\appendix

\section{The proof of properties of measure of $(k+1)$-partite entanglement }

It's easy to verify by definition of $(k+1)$-PE concurrence  $E_{k}(\rho)$ that item (P1) is true for all $k$-producible state, and item (P2) holds for all quantum states containing $(k+1)$-partite entanglement.
For any  subset $B$ of $\{1,2,\cdots,N\}$,  $\textrm{Tr}(\rho^2_{B})=\textrm{Tr}(\tilde{\rho}^2_{B})$, where $\tilde{\rho}_B$ is reduced density matrix of subsystem $B$ with $\tilde{\rho}=U_1\otimes\cdots\otimes U_N\rho U_1^\dag\otimes\cdots\otimes U_N^\dag$, then  item (P3) is valid.

In order to prove  item (P4), let  $\sum\limits_ip_i\rho_i=\rho$,
and suppose that $E_k(\rho_i)=\sum\limits_{j}q_{ij}E_k(|\varphi_{ij}\rangle\langle\varphi_{ij}|)$ with
$E_k(\rho_i)$ being given under the pure state decompositions $\{q_{ij},|\varphi_{ij}\rangle\}$ of the  state $\rho_i$.
Based on the above, we have $E_k(\rho)=E_k(\sum\limits_ip_i\rho_i)$, $\rho=\sum\limits_ip_i\rho_i=\sum\limits_{i,j}p_iq_{ij}|\varphi_{ij}\rangle\langle\varphi_{ij}|$.
Then we can obtain
\begin{equation}\label{}\nonumber
E_k(\sum\limits_ip_i\rho_i)=E_k(\rho)
\leq\sum\limits_{i,j}p_iq_{ij}E_k(|\varphi_{ij}\rangle\langle\varphi_{ij}|)=\sum\limits_{i}p_i\big[\sum\limits_{j}q_{ij}E_k(|\varphi_{ij}\rangle\langle\varphi_{ij}|)\big]
=\sum\limits_{i}p_iE_k(\rho_i),
\end{equation}
where  the inequality is valid because of the definition of $(k+1)$-PE concurrence. So we've proved that $E_{k}(\rho)$ is convex.

 We first prove that item (P5) holds for the pure state $|\psi\rangle$. For any the partitions $A_1|\cdots|A_m$ satisfying (\ref{partition}),
if we think of $|\psi\rangle$ as a bipartite quantum state of $\mathcal{H}_{A_t}\otimes
\mathcal{H}_{\bar{A_t}}, $
the concurrence of bipartite quantum pure states  $C_{A_t|\bar{A_t}}(|\psi\rangle)=\sqrt{2[1-\textrm{Tr}(\rho^2_{A_t})]}$ is nonincreasing under LOCC \cite{HWprl97,Rungta2001,MintertKus2005,Chen2005}, that is, $C_{A_t|\bar{A_t}}(\Lambda_{\textrm{LOCC}}(|\psi\rangle))\leq C_{A_t|\bar{A_t}}(|\psi\rangle)$. Then one has
\begin{equation}\label{}\nonumber
E_{k}(\Lambda_{\textrm{LOCC}}(|\psi\rangle))=\min\limits_{A}\dfrac{\sum\limits_{t=1}^{m}C_{A_t|\bar{A_t}}(\Lambda_{\textrm{LOCC}}(|\psi\rangle))}{m}
\leq\min\limits_{A}\dfrac{\sum\limits_{t=1}^{m}C_{A_t|\bar{A_t}}(|\psi\rangle)}{m}=E_{k}(|\psi\rangle).
\end{equation}
According to the convexity of $E_k(\rho)$ and the fact that $E_k(\rho)$ does not increase under LOCC for pure states, we can easily conclude that it does not increase under LOCC for mixed states, and thus item (P5) holds.

Let's  check that item (P6) is true for both $\rho$ and $\sigma$ are pure states.
Suppose that there exit partitions  $A_1|\cdots|A_m$  and $B_1|\cdots|B_n$ satisfying (\ref{partition}) such that
\begin{equation}\label{}\nonumber
E_{k}(\rho)=\dfrac{\sum\limits_{t=1}^{m}\sqrt{2\left[1-\textrm{Tr}(\rho^2_{A_t})\right]}}{m},
E_{k}(\sigma)=\dfrac{\sum\limits_{t=1}^{n}\sqrt{2\left[1-\textrm{Tr}(\sigma^2_{B_t})\right]}}{n},
\end{equation}
respectively. Based on these and the definition of $(k+1)$-PE concurrence for pure states, we have
\begin{equation}\label{}\nonumber
\begin{array}{rl}
E_{k}(\rho\otimes\sigma)
\leq&\dfrac{\sum\limits_{t=1}^{m}\sqrt{2\left[1-\textrm{Tr}(\rho^2_{A_t})\right]}+\sum\limits_{t=1}^{n}\sqrt{2\left[1-\textrm{Tr}(\sigma^2_{B_t})\right]}}{m+n}\\
\leq&\dfrac{\sum\limits_{t=1}^{m}\sqrt{2\left[1-\textrm{Tr}(\rho^2_{A_t})\right]}}{m}+\dfrac{\sum\limits_{t=1}^{n}\sqrt{2\left[1-\textrm{Tr}(\sigma^2_{B_t})\right]}}{n}\\
=&E_{k}(\rho)+E_{k}(\sigma).
\end{array}
\end{equation}
So, $(k+1)$-PE concurrence owns  item (P6) when  both $\rho$ and $\sigma$ are pure states.
Using the the definition of $(k+1)$-PE concurrence for mixed states,  convexity of $(k+1)$-PE concurrence and the effectiveness of item (P6) for  both $\rho$ and $\sigma$ are pure states,
 we can also deduce that the item (P6) holds  when $\rho$ and $\sigma$ are mixed states.


\begin{thebibliography}{99}

\bibitem{RMP81.865} R. Horodecki, P. Horodecki, M. Horodecki, and K. Horodecki, Quantum entanglement, \href{https://doi.org/10.1103/RevModPhys.81.865}{Rev. Mod. Phys. \textbf{81}, 865 (2009).}

\bibitem{Nature2000} C. H. Bennett and D. P. DiVincenzo, Quantum information and computation, \href{https://doi.org/10.1038/35005001}{Nature \textbf{404}, 247 (2000).}

\bibitem{PRL70.1895} C. H. Bennett, G. Brassard, C. Cr\'{e}peau, R. Jozsa, A. Peres, and W. K. Wootters, Teleporting an unknown quantum state via dual classical and Einstein-Podolsky-Rosen channels, \href{https://doi.org/10.1103/PhysRevLett.70.1895}{Phys. Rev. Lett. \textbf{70}, 1895 (1993).}

\bibitem{EPL84.50001} T. Gao, F. L. Yan, and Y. C. Li, Optimal controlled teleportation, \href{https://doi.org/10.1209/0295-5075/84/50001}{Europhys. Lett. \textbf{84}, 50001 (2008).}

\bibitem{PRL69.2881} C. H. Bennett and S. J. Wiesner,  Communication via one- and two-particle operators on Einstein-Podolsky-Rosen states, \href{https://doi.org/10.1103/PhysRevLett.69.2881}{Phys. Rev. Lett. \textbf{69}, 2881 (1992).}

\bibitem{QIC2007} M. B. Plenio and S. Virmani, An introduction to entanglement measures, Quantum Inf. Comput. \textbf{7}, 1 (2007).

\bibitem{PR474.1} O. G\"{u}hne and G. T\'{o}th, Entanglement detection, \href{https://doi.org/10.1016/j.physrep.2009.02.004}{Phys. Rep. \textbf{474}, 1 (2009).}

\bibitem{JPA2014} C. Eltschka and J. Siewert, Quantifying entanglement resources, \href{https://doi.org/10.1088/1751-8113/47/42/424005}{J. Phys. A \textbf{47}, 424005 (2014).}

\bibitem{TCS2002} B. M. Terhal, Detecting quantum entanglement, \href{https://doi.org/10.1016/S0304-3975(02)00139-1}{Theor. Comput. Sci. \textbf{287}, 313 (2002).}

\bibitem{AMP2010} M. Li, S. M. Fei, and X. Q. Li-Jost, Quantum entanglement: separability, measure,
fidelity of teleportation, and distillation, \href{https://doi.org/10.1155/2010/301072}{Adv. Math. Phys. \textbf{2010}, 301072 (2010).}

\bibitem{FP2013} M. Li, M. J. Zhao, S. M. Fei, and Z. X. Wang, Experimental detection of quantum entanglement, \href{https://doi.org/10.1007/s11467-013-0355-3}{Front. Phys. \textbf{8}, 357 (2013).}

\bibitem{NRP2019} N. Friis, G. Vitagliano, M. Malik, and M. Huber, Entanglement certification from theory to experiment, \href{https://doi.org/10.1038/S42254-018-0003-5}{Nat. Rev. Phys. \textbf{1}, 72 (2019).}

\bibitem{EPL104.20007} T. Gao, Y. Hong, Y. Lu, and F. L. Yan, Efficient $k$-separability criteria for mixed multipartite quantum states, \href{http://doi.org/10.1209/0295-5075/104/20007}{Europhys. Lett. \textbf{104}, 20007 (2013).}

\bibitem{QIC2008} A. S. M. Hassan and P. S. Joag, Separability criterion for multipartite quantum states based on the Bloch representation of density matrices, Quantum Inf. Comput. \textbf{8}, 773 (2008).

\bibitem{QIC2010} A. Gabriel, B. C. Hiesmayr, and M. Huber, Criterion for $k$-separability in mixed multipartite systems, Quantum Inf. Comput. \textbf{10}, 829 (2010).

\bibitem{PRA82.062113} T. Gao and  Y. Hong, Detection of genuinely entangled and nonseparable $n$-partite quantum states, \href{https://doi.org/10.1103/PhysRevA.82.062113}{Phys. Rev. A \textbf{82}, 062113 (2010).}



\bibitem{PRA91.042313} Y. Hong, S. Luo, and H. Song, Detecting $k$-nonseparability via quantum Fisher information, \href{https://doi.org/10.1103/PhysRevA.91.042313}{Phys. Rev. A \textbf{91}, 042313 (2015).}

\bibitem{SR5.13138} L. Liu, T. Gao, and F. L. Yan, Separability criteria via sets of mutually unbiased measurements, \href{https://doi.org/10.1038/srep13138}{Sci. Rep. \textbf{5}, 13138 (2015).}

\bibitem{PRA93.042310} Y. Hong and S. Luo, Detecting $k$-nonseparability via local uncertainty relations, \href{https://doi.org/10.1103/PhysRevA.93.042310}{Phys. Rev. A \textbf{93}, 042310 (2016).}

\bibitem{SciChina2017} L. Liu, T. Gao, and F. L. Yan,  Separability criteria via some classes of measurements, \href{https://doi.org/10.1007/S11433-017-9070-4}{Sci. China Phys. Mech. Astron. \textbf{60}, 100311 (2017).}

\bibitem{CPB2018} L. Liu, T. Gao, and F. L. Yan, Detecting high-dimensional multipartite entanglement via some classes of measurements, \href{https://doi.org/10.1088/1674-1056/27/2/020306}{Chinese Phys. B \textbf{27}, 020306 (2018).}

\bibitem{QIP2020} W. Xu, C. J. Zhu, Z. J. Zheng, and S. M. Fei, Necessary conditions for classifying $m$-separability of multipartite entanglements, \href{https://doi.org/10.1007/s11128-020-02705-6}{Quantum Inf. Process. \textbf{19}, 200 (2020).}


\bibitem{PRA68.042307} T. C. Wei and P. M. Goldbart, Geometric measure of entanglement and applications to bipartite and multipartite quantum states, \href{https://dx.doi.org/10.1103/PhysRevA.68.042307}{Phys. Rev. A \textbf{68}, 042307 (2003).}

\bibitem{PRL93.230501} A. R. R. Carvalho, F. Mintert, and A. Buchleitner, Decoherence and multipartite entanglement, \href{https://doi.org/10.1103/PhysRevLett.93.230501}{Phys. Rev. Lett. \textbf{93}, 230501 (2004).}

\bibitem{PRA83.062325} Z. H. Ma, Z. H. Chen, J. L. Chen, C. Spengler, A. Gabriel, and M. Huber, Measure of genuine multipartite
entanglement with computable lower bounds, \href{https://dx.doi.org/10.1103/PhysRevA.83.062325}{Phys. Rev. A \textbf{83}, 062325 (2011).}

\bibitem{PRA86.062323} Y. Hong,  T. Gao, and F. L. Yan,  Measure of multipartite entanglement with computable lower bounds, \href{https://dx.doi.org/10.1103/PhysRevA.86.062323}{Phys. Rev. A \textbf{86}, 062323 (2012).}

\bibitem{PRL112.180501} T. Gao, F. L. Yan, and S. J. van Enk, Permutationally invariant part of a density matrix and nonseparability
of $N$-qubit states, \href{https://doi.org/10.1103/PhysRevLett.112.180501}{Phys. Rev. Lett. \textbf{112}, 180501 (2014).}


%
%
%



\bibitem{PLA2021} Y. Hong, T. Gao,  and F. L. Yan, Detection of $k$-partite entanglement and $k$-nonseparability of multipartite quantum states, \href{https://doi.org/10.1016/j.physleta.2021.127347}{Phys. Lett. A \textbf{401},  127347 (2021).}



\bibitem{HWprl97} S. Hill and W. K. Wootters, Entanglement of a pair of quantum bits,  \href{https://doi.org/10.1103/PhysRevLett.78.5022}{Phys. Rev. Lett. \textbf{78}, 5022 (1997).}

\bibitem{Rungta2001}P. Rungta, V. Bu\v{z}ek, C. M. Caves, M. Hillery, and G. J. Milburn, Universal state inversion and concurrence in arbitrary dimensions, \href{https://dx.doi.org/10.1103/PhysRevA.64.042315}{Phys. Rev. A \textbf{64}, 042315 (2001).}

\bibitem{MintertKus2005}F. Mintert, M. Kus, and A. Buchleitner, Concurrence of mixed multipartite quantum states, \href{https://doi.org/10.1103/PhysRevLett.95.260502}{Phys. Rev. Lett. \textbf{95}, 260502 (2005).}

\bibitem{Chen2005} K. Chen, S. Albeverio, S. M. Fei, Concurrence of arbitrary dimensional bipartite quantum states, \href{https://doi.org/10.1103/PhysRevLett.95.040504}{Phys. Rev. Lett. \textbf{95}, 040504  (2005).}


\end{thebibliography}
\end{document}